\input{aipcheck}

\documentclass[
    ,final            
  ]
  {aipproc}

\layoutstyle{6x9}
\usepackage{amsmath}
\begin{document}

\title{Shortest movie: Bose-Einstein Correlation functions in $\mathrm{e}^+\mathrm{e}^-$
annihilations}



\author{Tam\'as Nov\'ak \\ 
(L3 Collaboration)}
{
  address={Radboud University Nijmegen/NIKHEF, Nijmegen, The Netherlands}
}

\begin{abstract}
Bose-Einstein correlations of identical charged-pion pairs
produced in hadronic Z decays are analyzed in terms of various parametrizations.
A good description is achieved using L\'evy stable distributions. 
The source function is reconstructed with the help of the $\tau$-model.
\end{abstract}

\maketitle


\section{Introduction}
In particle and nuclear physics intensity interferometry provides a direct
experimental method for the determination of sizes, shapes and lifetimes
of particle-emitting sources (for recent reviews see \cite{Wolfram,Tamas1}).
In particular, boson interferometry provides a powerful tool for the 
investigation of the space-time structure of particle production processes, 
since Bose-Einstein correlations (BEC) of two identical bosons reflect both 
geometrical and dynamical properties of the particle radiating source. 

In $\mathrm{e}^+\mathrm{e}^-$ annihilation BEC are maximal if the invariant
momentum difference is small, even when one of the relative momentum components
is large, as was seen by TASSO \cite{tasso} and which we have confirmed. 
For a hydrodynamical type of source, 
on the contrary, BEC decrease when any of the relative momentum components is large 
\cite{Tamas1, Tamas2}. 

Here we investigate various parametrizations and find that a good description
of the Bose-Einstein correlation function can be achieved using L\'evy stable 
distributions as the source function. Within the framework of models assuming 
strongly correlated coordinate and momentum space, we then reconstruct the complete 
space-time picture of the particle emitting source in hadronic Z decay.

For our analysis we use a sample of about 500 thousand two-jet events, 
selected by the Durham algorithm \cite{durham} with $y_{\mathrm{cut}}=0.006$,
from e$^+$e$^-$annihilation data collected by L3 at a center-of-mass energy of 91.2 GeV. 
\section{Parametrizations of BEC}
The two-particle correlation function is defined as:
\begin{equation}
  R_2(p_1,p_2) = \frac{\rho_2(p_1,p_2)}
    {\rho_1(p_1) \rho_1(p_2)},
\end{equation}
where $\rho_2(p_1,p_2)$ is the two-particle invariant momentum 
distribution, $\rho_1(p_i)$ the single-particle invariant momentum 
distributions and $p_i$ the four-momentum of particle $i$. Since we are only 
interested in BEC, the product of single particles densities is replaced by the
so-called reference sample, 
$\rho_0(p_1,p_2)$, the two-particle density that would occur in the absence of Bose-Einstein
interference. Here we use mixed events as a reference sample.

After some assumptions \cite{Wolfram,Tamas1}, this two-particle correlation function 
is related to the 
Fourier transformed source distribution. In this case
\begin{equation}
  R_2(p_1,p_2) = 1 + |\tilde{f} (Q)|^2 ,
\end{equation}
where $f(x)$ is the density distribution of the source,
$Q$ is the invariant four-momentum difference, $Q=-(p_1-p_2)^2$ and
$\tilde{f} (Q)$ is the Fourier transform of $f(x)$.

\subsection{Gaussian distributed source}
The simplest assumption is that the source has a symmetric Gaussian distribution,
in which case  
$\tilde{f} (Q)=\exp \left(i\mu Q -\frac{(RQ)^2}{2}\right)$ and
\begin{equation}\label{gauss}
  R_2(Q) = \gamma \left[1 + \lambda \exp \left(-(R Q)^2 \right) \right] 
           \left( 1 + \delta Q \right),
\end{equation}
where the parameter $\gamma$ is a constant of normalization,
$\lambda$ an incoherence factor, which measures the strength of
the correlation, and $\left( 1 + \delta Q \right)$ is introduced to parametrize
possible long-range correlations not adequately accounted for in the reference sample.

A fit of Eq.(\ref{gauss}) to the data results in an unacceptably low confidence
level from which we conclude that the shape of the source deviates from a 
Gaussian. The fit is particularly bad at low $Q$ values.

\subsection{L\'evy distributed source}
Adopting Nolan's $S(\alpha, \beta=0, \gamma, \delta; 1)$ convention \cite{Nolan} for the symmetric 
L\'evy stable distribution with rescaling of the scale parameter $\gamma$ to $R$ and
the location parameter $\delta$ to $x_0$, the Fourier transform (characteristic function)
$\tilde{f}(Q)$ has the following general form:
\begin{equation}
  \tilde{f}(Q) = \exp (iQ x_0 - |R Q|^\alpha).
\end{equation}
The index of stability, $\alpha$, satisfies the inequality $0<\alpha \leq 2$. 
The case $\alpha=2$ corresponds to a Gaussian source distribution. For more details see 
\cite{Nolan}.

Then $R_2$ has the following, relatively simple form \cite{Tamas3}:
\begin{equation}\label{symlev}
  R_2(Q) = \gamma \left[ 1+ \lambda \exp \left(-(RQ)^\alpha \right) \right]
          (1+ \delta Q).
\end{equation}
After fitting  Eq.(\ref{symlev}) to the data it is clear that the correlation
function is far from Gaussian: $\alpha \approx 1.3$. The confidence level, although 
improved compared to the fit of Eq.(\ref{gauss}), is still unacceptably low.

Since there is no particle production before the
onset of the collision, a more appropriate form of the source distribution
for the time component is the  asymmetric stable distribution. In this 
case, one obtains the following result for the correlation function \cite{Tamasbesz}:
\begin{equation}\label{asymlev}
  R_2(Q) = \gamma \left[ 1+ \lambda \cos \left[(R_\mathrm{a}Q)^ \alpha \right]
           \exp \left(-(RQ)^\alpha \right) \right] (1+ \delta Q),
\end{equation}
where $R_\mathrm{a}$ is an additional parameter, a measure of the onset of particle production
\cite{Nolan, Tamas3}.

\begin{figure}\label{fig2}
  \includegraphics[height=.31\textheight]{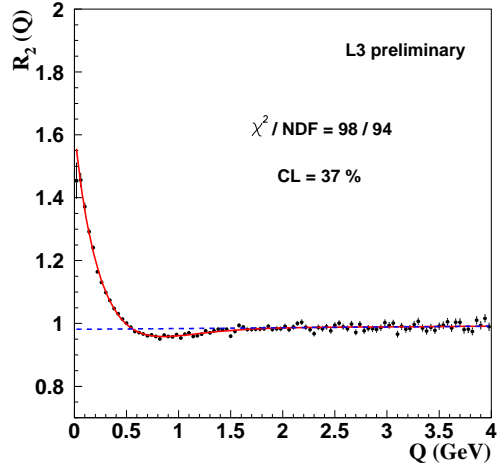}
  \caption{The Bose-Einstein correlation function $R_2$. The curve 
corresponds to the fit of Eq.(\ref{asymlev}).}
\end{figure}

The fit of Eq.(\ref{asymlev}) to the data, shown in Figure \ref{fig2}, is statistically 
acceptable. The data are well described by the fit. Note that for $Q$ between 
0.5 $\mathrm{GeV}$ and 1.5 $\mathrm{GeV}$ the data points go below the dashed line,
which stands for the long-range correlations extrapolated to lower $Q$ values.
These data points indicate an anti-correlation in the $Q \approx 1$
GeV region.  This property of the data is well reproduced by the fitted curve, which
also goes below unity as a result of
the cosine term in Eq.(\ref{asymlev}), which comes from the asymmetric 
L\'evy assumption. The fitted value of $\alpha$ is $0.82 \pm 0.03$.


\section{The $\tau$-model}
A model of strongly correlated phase-space was developed \cite{Tamas4} to 
explain the experimentally found invariant relative momentum $Q$
dependence of Bose-Einstein correlations in $\mathrm{e}^+\mathrm{e}^-$ reactions.
This model also predicts a specific transverse mass dependence of $R_2$,
that we subject to an experimental test here.

In this model, it is assumed that the average production point $\overline{x}^\mu$ of
particles with a given momentum $k^\mu$ is given by
\begin{equation}
  \overline{x}^\mu (k^\mu)  = d k^\mu.
\end{equation}
In the case of two-jet events, $d = \frac{\tau}{m_\mathrm{T}}$, where 
$\tau = \sqrt{t^2 - k_z^2}$ is the longitudinal proper-time and
$m_\mathrm{T}=\sqrt{m^2+p_\mathrm{T}^2}$ is the transverse mass. 
The second assumption is that the distribution of $x^\mu (k^\mu)$ about its average,
$\delta_\Delta ( x^\mu(k^\mu) - \overline{x}^\mu (k^\mu) )$, is narrower than the 
proper-time distribution.
Then the emission function of the $\tau$-model is
\begin{equation} \label{source}
  S(x,k) = \int_0^{\infty} \mathrm{d}\tau H(\tau)\delta_{\Delta}(x-dk) N_1(k),
\end{equation}
where $H(\tau)$ is the longitudinal proper-time distribution, the factor 
$\delta_{\Delta}(x-dk)$ describes the strength of the correlations between 
coordinate space and momentum space variables and $N_1(k)$ is the experimentaly 
measurable single-particle spectrum.

In the plane-wave approximation the Yano-Koonin formula \cite{Yano} gives the following
two-pion multiplicity distribution:
\begin{equation}
   \rho_2(k_1,k_2) = \int \mathrm{d}^4 x_1 \mathrm{d}^4 x_2 S(x_1,k_1) S(x_2,k_2)
   \left( 1+ \cos \left[ (k_1-k_2) (x_1-x_2) \right] \right). 
\end{equation}
Approximating the $\delta_\Delta$ function by a Dirac delta function, 
the argument of the cosine becomes
\begin{equation}
 (k_1 - k_2)(\bar{x}_1 - \bar{x}_2) = - 0.5 (d_1 + d_2) Q^2. 
\end{equation}
Then the two-particle Bose-Einstein correlation function is approximated by
\begin{equation}
  R_2(k_1,k_2) = 1 + \lambda \mathrm{Re} \widetilde{H}^2 
\left(  \frac{Q^2}{2 \overline{m}_{\mathrm{T}}} \right),
\end{equation}
where $\widetilde{H} (\omega) = \int \mathrm{d} \tau H(\tau) \exp(i \omega \tau)$ 
is the Fourier transform of $H(\tau)$. Thus an invariant relative momentum dependent
BEC appears.

Guided by the result of the previous section,
we use an asymmetric L\'evy distribution for the longitudinal proper-time density.
Thus the corresponding BEC function has an analytic, 
although somewhat complicted form \cite{Tamas3,Tamasbesz}:
\begin{equation}
   R_2(Q^2,\overline{m}_\mathrm{T}) = 
       \gamma \Bigg[ 1+\lambda \cos \left( \frac{\tau_0 Q^2}{\overline{m}_\mathrm{T}} 
       +A  \left( \frac{\Delta \tau Q^2}{\overline{m}_\mathrm{T}} \right)^ \frac{\alpha}{2} \right) 
        \exp \left( -\left( \frac{\Delta\tau Q^2}{\overline{m}_\mathrm{T}} 
	  \right)^\frac{\alpha}{2} \right) \Bigg]   B
\end{equation}
where the parameter $\tau_0$ is the proper-time of the onset of particle
production, $\Delta \tau$ is a measure of the width of the proper-time
distribution, $A = \tan \left( \frac{\alpha \pi}{4} \right) $ and $B =(1+\delta Q)$ 

After fitting for various $\overline{m}_\mathrm{T}$ interval we find that the 
quality of the fits 
is statistically acceptable and the fitted values of the model parameters are stable
and within errors the same in all investigated $m_{\mathrm{T}}$ interval.
The $\tau$-model with a one-sided Levy proper-time distribution describes the data with
parameters $\tau_0= 0$ fm, $\alpha \approx 0.80\pm 0.05$ and $\Delta \tau \approx 2.0\pm0.6$ fm.
\section{Reconstruction of the emission function}

In order to reconstruct the space-time picture of the emitting process we assume that
the emission function can be factorized in the following way:
\begin{equation}
  \label{eq:fact}
  S(r,z,t) = I(r) G(\eta) H(\tau),
\end{equation}
where $I(r)$ is the single-particle transverse distribution, $G(\eta)$ is
the space-time rapidity distribution of particle production,
which approximately coincides with the single-particle rapidity distribution, 
and $H(\tau)$ is the observed proper-time distribution.
\begin{figure}\label{longemis}
  \includegraphics[height=.25\textheight,clip=]{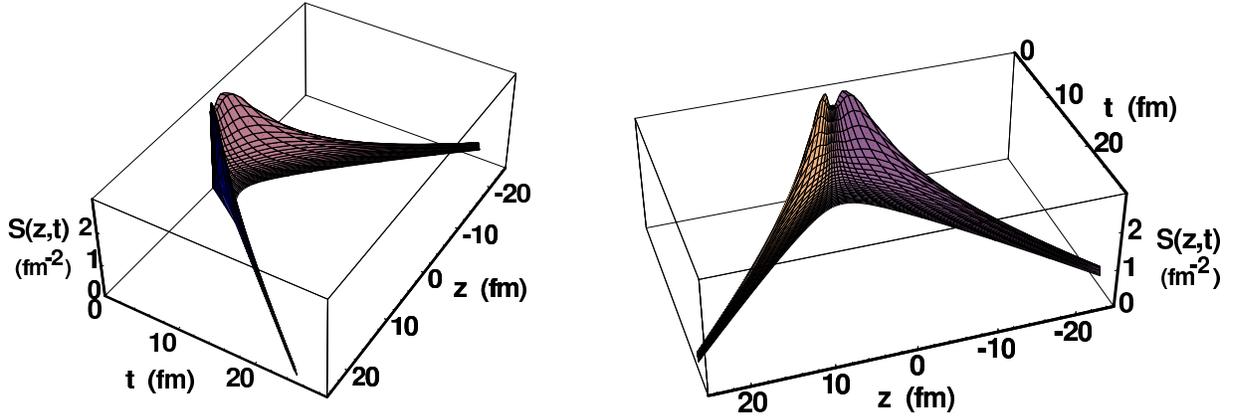}
  \caption{Two views of the longitudinal part of the source function normalized to the 
average number of pions per event.}
\end{figure}

With these assumptions one can reconstruct the longitudinal part of the emission
function integrated over the transverse distribution. It is plotted as a function of
$t$ and $z$ in Figure \ref{longemis}. It exhibits the typical boomerang shape with a 
maximum at low $t$ and $z$ but with tails reaching out to very large $t$ and $z$ 
values, a feature already observed for hadron-hadron \cite{NA22} and 
heavy ion collisions\cite{Ster}.

\begin{figure}\label{movie}
  \includegraphics[height=.61\textheight,clip=]{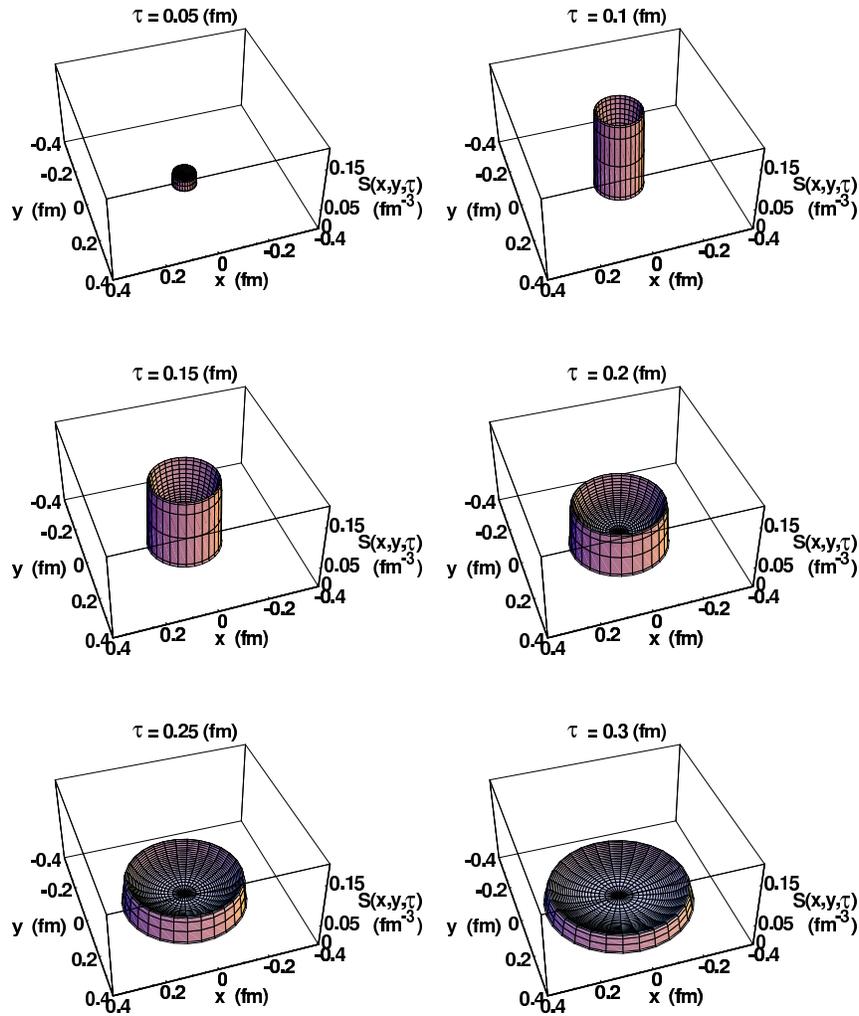}
  \caption{The source function normalized to the average number of pions per event for 
various proper-times.}
\end{figure}

The transverse profile, which follows from Eq. (\ref{source}), has the following form:
\begin{equation}
   \frac{\mathrm{d}^4 n}{\mathrm{d}\tau \mathrm{d}^3 r} = \frac{m_\mathrm{T}^3}{\tau ^3}
   H(\tau) N_1\left( k=\frac{m_\mathrm{T} r}{\tau} \right) .
\end{equation}
This equation describes the particle production in coordinate space as a function of the
proper-time $\tau$. It describes the expansion of the source as the proper-time increases.
The particle production probability is proportional to the proper-time distribution 
$H(\tau)$. Figure \ref{movie} shows the transverse part of the emission function for 
various proper-times. Particle production starts immediately, increases rapidly and decreases
slowly. A ring-like structure, similar to the expanding, ring-like wave created by a pebble in a pond,
is reconstructed from L3 data, as shown in Fig. 3. An animated gif file that shows this effect 
is available from \cite{Movie}.
 
\begin{theacknowledgments}
The author would like to thank T. Cs\"org\H{o}, W. Kittel and W. Metzger
for inspiration, support and careful attention, as well as to all 
members of the L3 collaboration.
\end{theacknowledgments}

\end{document}